\documentclass[twocolumn,pre,showpacs,preprintnumbers,amsmath,amssymb,floatfix,superscriptaddress]{revtex4}
\usepackage{graphicx}
\usepackage{amsmath}
\usepackage{amssymb}
\usepackage{color}

\begin{document}

\title{Percolative properties of hard oblate ellipsoids of revolution with a soft shell}

\author{Gianluca Ambrosetti}\email{gianluca.ambrosetti@epfl.ch}\affiliation{LPM, Ecole Polytechnique F\'ed\'erale de Lausanne, Station 17, CH-1015 Lausanne, Switzerland}\affiliation{ICIMSI, University of Applied Sciences of
Southern Switzerland, CH-6928 Manno, Switzerland}
\author{Niklaus Johner}\affiliation{LPM, Ecole Polytechnique F\'ed\'erale de
Lausanne, Station 17, CH-1015 Lausanne, Switzerland}
\author{Claudio Grimaldi}\affiliation{LPM, Ecole Polytechnique F\'ed\'erale de
Lausanne, Station 17, CH-1015 Lausanne, Switzerland}
\author{Andrea Danani}\affiliation{ICIMSI, University of Applied Sciences of
Southern Switzerland, CH-6928 Manno, Switzerland}
\author{Peter Ryser}\affiliation{LPM, Ecole Polytechnique F\'ed\'erale de
Lausanne, Station 17, CH-1015 Lausanne, Switzerland}



\begin{abstract}
We present an in-depth analysis of the geometrical percolation
behavior in the continuum of random assemblies of hard oblate
ellipsoids of revolution. Simulations where carried out by
considering a broad range of aspect-ratios, from spheres up to
aspect-ratio 100  plate-like objects, and with various limiting
two particle interaction distances, from 0.05 times the major axis
up to 4.0 times the major axis. We confirm the widely reported
trend of a consistent lowering of the hard particle critical
volume fraction with the increase of the aspect-ratio. Moreover,
assimilating the limiting interaction distance to a shell of
constant thickness surrounding the ellipsoids, we propose a
simple relation based on the total excluded volume of these
objects which allows to estimate the critical concentration  from
a quantity which is quasi-invariant over a large spectrum of
limiting interaction distances. Excluded volume and volume
quantities are derived explicitly.
\end{abstract}
\maketitle

\section{Introduction}

A central problem in materials science is the precise evaluation
of the percolation threshold of random particle dispersions
embedded in a continuous medium. This occurs typically in
composite materials and is of importance for the prediction of
relevant properties such as the electrical conduction in
insulator-conducting composites. Practical examples include
carbonaceous fillers like carbon fibers, graphite, carbon black,
carbon nanotubes and fullerenes, but also metallic and ceramic
ones, while matrices can be for instance polymeric, metallic or
ceramic. Despite that the most studied particle form is the
sphere, see e.g.
\cite{Heyes2006,Lee1998,Wang1993,Balberg1987,Shante1971,Scher1970},
a broad range of fillers in real composites have forms which
deviate consistently from the sphere. Previous investigations
have considered different particle shapes like e.g. sticks
\cite{Neda1999,Bug1986,Bug1985,Balberg1984a,Balberg1984b},wavy
sticks \cite{Berhan2007b}, plates
\cite{Celzard1996,Charlaix1986,Charlaix1984} or ellipsoids
\cite{Yi2004,Garboczi1995,Sevick1988,Skal1974}, but always in the
fully penetrable case, where the particles are allowed to freely
overlap. Only in few cases were hard sticks with a soft shell
considered \cite{Schilling2007,Berhan2007a,Ogale1993,Balberg1987}, and a recent
paper \cite{Akagawa2007} contemplated, as in the present study,
the case of hard ellipsoids of revolution, but in the prolate
domain.

The widespread use of composites containing fibrous fillers has
made the stick, or other elongated objects, the favorite
non-spherical shape in many studies. Nevertheless, some other
fillers, notably graphite, have shapes which are better
assimilable to flattened ellipsoids or platelets, and over a
broad range of aspect-ratios i.e. longer dimension to shorter
dimension ratios. The exploration of the relatively uncharted
terrain of the percolative properties of oblate objects as a
function of their aspect-ratio is then the aim of the present
study.

In this paper we consider the special case of oblate \emph{ellipsoids of revolution},
usually called (oblate) \emph{spheroids}, which are ellipsoids with two equal (major)
axes and may be obtained by rotation of a 2D ellipse around its minor axis. The reasons
for this choice are twofold: first, spheroids are characterized by a smaller number of
parameters (7 against 9 of the general ellipsoid); second, experimental measurement
techniques of the filler particle size distributions are generally able to extract only
major and minor dimensions, making it difficult to quantitatively define a size distribution
for the third axis.

Our model is defined by a dispersion of impenetrable spheroids of
identical dimensions with isotropic distribution of the symmetry
axis orientation. Given any two spheroids, a connectivity
criterion is introduced by allowing an upper cutoff distance
beyond which the two spheroids are considered disconnected. More
precisely, each spheroid is coated with a penetrable shell of
constant thickness, and two particles are connected if their
shells overlap. In a system of conducting spheroids dispersed in
an insulating continuum host, the shell thickness can be
physically interpreted as a typical tunneling length between the
particles, governing the electrical connectivity of the composite.

To carry out our investigation we exploit a simulation algorithm,
described in the following section, that allows to determine the
percolation behavior of a random distribution of impenetrable spheroids
as a function of their volume fraction, aspect-ratio, shell thickness
and simulation cell size.

\section{The Simulation Algorithm}
Even if the ellipsoid is a sufficiently simple geometrical form,
much less simple is the construction of an algorithm which
involves random distributions of them, since the computation of
relevant functions like the particle inter-distance or the overlap
criterion is far from trivial, as opposed to the sphere. Now, if
we want to build an algorithm to carry out the proposed
simulations, we first need a routine that generates a
distribution of ellipsoids and that calculates the inter-distance
between them. In order to do that we require two functions, an
ellipsoid overlap criterion and the distance between two
ellipsoids, the first being needed of course only if it can be
computed in a time consistently shorter than the second. None of
these functions allows simple closed form solutions, but some
evaluation techniques are nevertheless available
\cite{Donev2005,Yi2004,Garboczi1995,Perram1985,Vieillard1972}. We
have chosen the approach proposed by Rimon and Boyd (R\&B)
\cite{Rimon1997,Rimon1992} which was used for an obstacle
collision detection procedure for robots, where short
computational times are essential. The R\&B technique allows two
key benefits: 1)  A quick estimation procedure of the distance
between two ellipsoids that uses standard computation routines
and that can be made sufficiently precise. 2)  An overlap
criterion between two ellipsoids as an intermediate result to the
inter-distance computation that can be calculated in about half
the time needed for the complete calculation. The computation is
based on a formula for the distance of a point from an ellipsoid
which reduces the problem to the calculation of the minimal
eigenvalue of an auxiliary matrix constructed from the
geometrical data.

\begin{figure}[h]
    \begin{center}
  \includegraphics[scale=0.5, clip='true']{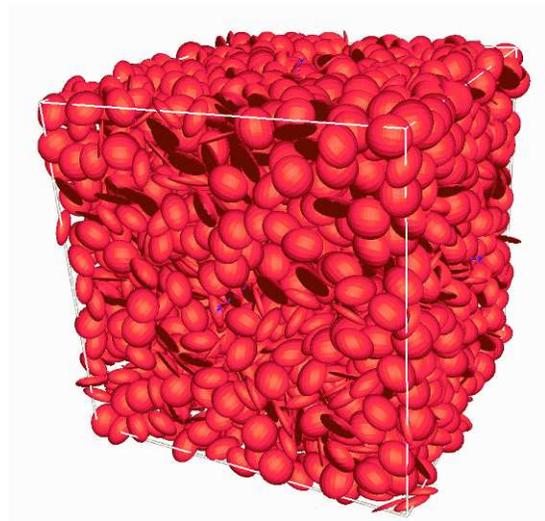}
  \caption{Distribution of $3000$ spheroids with
  aspect-ratio $a/b=10$.}\label{fig:SIM1}
  \end{center}
\end{figure}

We are now going to briefly outline how the distribution
algorithm is constructed. First, a spheroid distribution is
created inside a cubic cell of volume $L^3$ by random sequential
addition: for every new particle, random placing is attempted and
accepted as valid only if there is no overlap with any neighboring
particle. To speed up the search for neighbors the main cubic
cell is subdivided in discrete binning cells of size comparable
to the major dimension of the spheroids. Moreover, to avoid
unnecessary computations of the overlap function, cases are
discarded which unavoidably lead to overlap or that can anyway
not lead to overlap via simple geometrical rejection criteria.
Periodic boundary conditions are imposed on the main cell.
Second, the inter-particle distance is computed. An interaction
inter-distance is chosen so that spheroids separated by a
distance greater than it are considered non-interacting. Again,
the same neighbor search and rejection criteria are used and, if
it is the case, the distance computation is performed.  To do
this in an efficient way, a first R\&B calculation is executed
and if the resulting first distance estimate is clearly beyond the
interaction range even when considering the worst possible error
the calculation is stopped. Otherwise, the computation is
continued by performing the R\&B calculation with inverted
spheroids, comparing with the first calculation and retaining the
shorter of the two and finally reiterating part of the R\&B
procedure to obtain a further correction. This procedure leads to
a distance estimate that has an average error of about +1 \% on a
wide range of distance to major spheroid dimension ratios (from
$\textrm{10}^{\textrm{-4}}$ to 10), as obtained by comparing the
R\&B results with a more accurate but much slower distance
evaluation procedure. Figure \ref{fig:SIM1} shows one of such
distributions as it appears by loading the algorithm output file
in a viewer.

Once the desired distribution has been created and the neighboring particles
inter-distances computed, the distribution algorithm output data are fed into
the part of a program which isolates the connected cluster using a modified
version of the Hoshen-Kopelman algorithm \cite{Johner2008,Al-Futaisi2003,Hoshen1976}.
Finally, it is verified if the connected cluster spans two specific opposite sides of the simulation cell.

\section{Simulation Results}

To explore the percolative properties of hard oblate spheroids
surrounded with a penetrable shell of constant thickness, we
considered spheroids with an aspect-ratio, i.e., spheroid major
axis $a$ to minor (simmetry) axis $b$ ratio $a/b$ between $1$
(spheres) and $100$. The shell thickness $d$ to spheroid major
axis ratio $d/a$ was chosen to variate between $0.05$ and $4.0$.

To extrapolate the percolation threshold from the simulation
algorithm we followed finite-size scaling arguments as described
in Ref. \cite{Rintoul1997}, and briefly outlined below. For a
given size $L$ of the cube, we obtained the spanning probability
as a function of the spheroids volume fractions by recording the
number of times a percolating cluster appeared over a given number
of realizations. The resulting spanning probabilities were then
plotted against the volume fraction and fitted with the sigmoidal
function
\begin{equation}
\label{eq:sigmoid}
f=\frac{1}{2}\bigg[1+\tanh{\bigg(\frac{\phi-\phi^{\textrm{eff}}_{c}}{\Delta}\bigg)}\bigg]\quad,
\end{equation}
where $\phi^{\textrm{eff}}_{c}$ is the percolation threshold for
a given value of $L$ and corresponds to the hard particle volume
fraction at which the spanning probability is equal to
$\frac{1}{2}$, while $\Delta$ represents the width of the
percolation transition. Both $\phi^{\textrm{eff}}_{c}$ and
$\Delta$ depend on the size $L$ of the system and, by following
the scaling arguments of \cite{Rintoul1997}, allow to deduct the
percolation threshold $\phi_{c}$ for the infinite system through
the following scaling relations:
\begin{align}
\label{eq:scaling1}
&\Delta(L)\propto L^{-\frac{1}{\nu}},\\
\label{eq:scaling2} &\phi^{\textrm{eff}}_{c}(L)-\phi_{c}\propto
L^{-\frac{1}{\nu}},
\end{align}
where $\nu$ is the correlation length exponent. By repeating the
simulation procedure for different cell sizes it is possible, via
the percolation transition widths $\Delta$ and the inversion of
Eq. \eqref{eq:scaling1}, to extract $\nu$ and consequently, from
Eq. \eqref{eq:scaling2}, the percolation threshold $\phi_{c}$ for
$L=\infty$. We choose to simulate ten different cell sizes, $L=
10$, $13$, $15$, $17$, $20$, $23$, $25$, $27$, $30$ and $35$
times the major spheroid dimension, i.e., twice the major axis
$a$. For thick shells ($d/a\geq 1.0$) the cell sizes were
increased further. The spheroid number was in the order of
thousands for the smallest cells up to about $70'000$ for the
largest. The number of realizations per volume fraction step
varied from $50$ for the smallest shell thickness up to $400$ for
the thicker ones. Higher realization numbers did not show appreciable improvements. In all
cases, the correlation length exponent $\nu$ had a value around
0.9, in good agreement with previous results on spheres
\cite{Johner2008,Lee1998,Rintoul1997}. However, sometimes the
fluctuations of the $\phi^{\textrm{eff}}_{c}$ where too large and
a simple average of the results provided a more significative
result that the one obtained from the finite size analysis.

\begin{figure}[t!]
    \begin{center}
  \includegraphics[scale=0.5, clip='true']{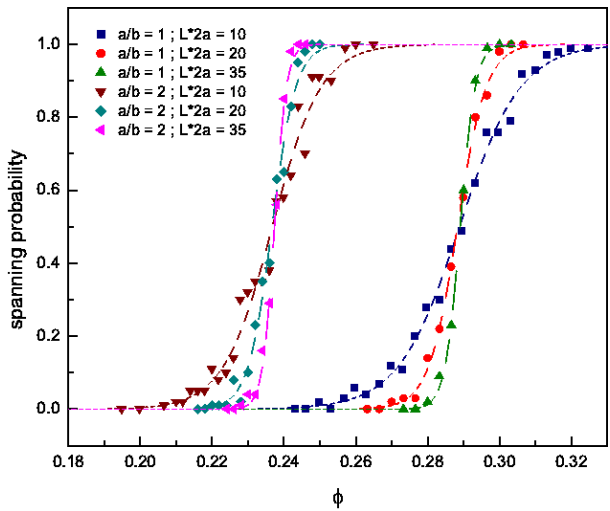}
  \caption{Percolation width variation with the increase of the
  simulation cell size for the aspect-ratio 1 and 2 cases.
  $d/a=0.1111$.}\label{fig:SIM2}
  \end{center}
\end{figure}

\begin{figure}[t!]
    \begin{center}
  \includegraphics[scale=0.9, clip='true']{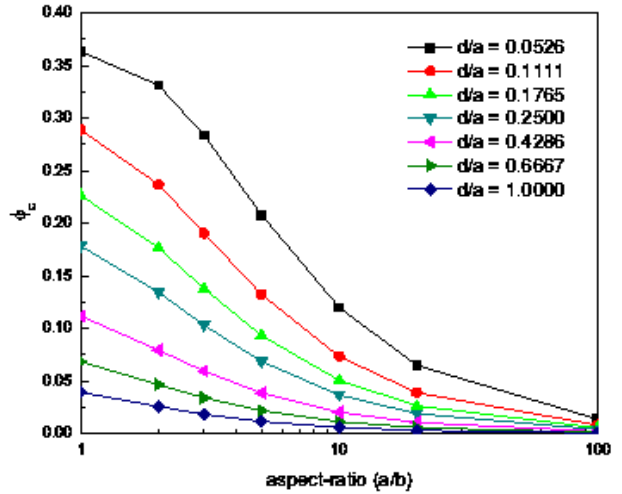}
  \caption{Percolation threshold $\phi_{c}$ variation a a function of
  the aspect-ratio for different shell thicknesses.}\label{fig:SIM02}
  \end{center}
\end{figure}

In Fig. \ref{fig:SIM2} we report the obtained spanning probability
as a function of $\phi$ for $a/b=1$ and $a/b=2$ and for selected
values of the cell size $L$. The shell thickness $d$ to major axis
ratio was set equal to $d/a=0.1111$. From the figure it is clear
that increasing the aspect-ratio from $a/b=1$ (spheres) to
$a/b=2$ leads to a lowering of the percolating volume fraction.
This trend is confirmed in Fig. \ref{fig:SIM02}, where the
critical hard particle volume fraction $\phi_{c}$ is plotted as a
function of $a/b$ and for several values of the penetrable shell
thickness. For the thinnest shells we find that $\phi_c$ can be
reduced by about one order of magnitude in going from $a/b=1$ up
to $a/b=100$. This result is fully consistent with the frequently
reported trend that assemblies of oblate objects with high
aspect-ratios entail a lower percolation threshold. For example,
several studies of graphite-polymer composites reported a
consistent lowering of the electrical conductivity percolation
threshold when very high aspect-ratio graphite nanosheets
\cite{Lu2006,Chen2003,Celzard1996} or graphene flakes
\cite{Stankovich2006} were used.

\begin{figure}[h!]
    \begin{center}
  \includegraphics[scale=0.5, clip='true']{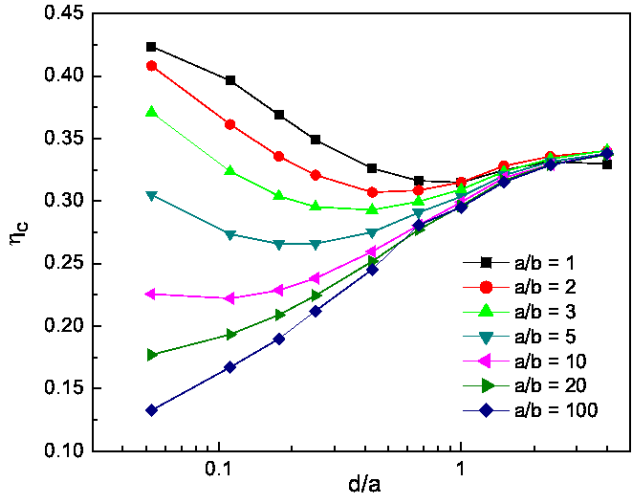}\\
  \caption{$\eta_{c}$ as a function of the shell thickness for different
  aspect-ratios.}\label{fig:SIM3}
  \end{center}
\end{figure}

Besides $\phi_c$, another quantity characterizing the percolation
threshold is the reduced critical density $\eta_c$ defined as
\begin{equation}
\eta_{c}= \rho_{c}V_{d}=\phi_{c}\frac{V_{d}}{V},
\end{equation}
where $\rho_c$ is the number density at percolation and $V_{d}$
is the total object volume, comprising the volume of the hard
spheroid, $V$, plus that of the penetrable shell. $V_d$ is
explicitly calculated in the appendix, see Eq. (\ref{eq:Vd}). The
behavior of $\eta_c$, plotted in Fig. \ref{fig:SIM3} as a function
of the penetrable shell thickness $d/a$ and for several
aspect-ratios, accounts for the dependence of the percolation
threshold on the geometry of the total object (hard-core plus
penetrable shell). Indeed, for $d/a=4$ the reduced critical
density is almost independent of the aspect-ratio $a/b$ while, for
thinner penetrable shells, $\eta_c$ acquires a stronger dependence
on $a/b$. This is due to the fact that, for large $d/a$ values,
the form of the total object does not deviate much from that of a
sphere, so that $\eta_c\simeq 0.34$ as for fully penetrable
spheres. On the contrary, for smaller values of $d/a$, the
geometry of the total object is more similar to that of an oblate
ellipsoid, with a consequently stronger dependence of $\eta_c$ on the
aspect-ratio.

\section{Quasi-invariants at the Percolation Threshold}
\label{sec:general SEE}

In continuum percolation, an important quantity providing information
on the local topology of the percolating cluster is the average number
$B_c$ of objects connected to a given particle. For fully penetrable objects,
and since in this case there is no spatial correlation, $B_c$ is simply given
by \cite{Balberg1984a}:
\begin{equation}
\label{eq:Bcspheres}
B_c=\rho_{c}V_{ex},
\end{equation}
where $V_{ex}$ is the excluded volume defined by the volume around an object
where the center of another object cannot penetrate if overlap is to be avoided.
For penetrable spheres each of volume $V$, the excluded volume is $V_{ex}=8V$ and,
by using $\rho_c=\eta_c/V$ with $\eta_c\simeq 0.34$, the resulting connectivity
number is $B_c\simeq 2.74$, which agrees well with the evaluation of $B_c$
from a direct enumeration of connections in assemblies of penetrable
spheres at percolation \cite{Balberg1987,Heyes2006}.
Indeed, for fully penetrable spheres, for which the sphere centers are distributed randomly,
Eq. \eqref{eq:Bcspheres} simply states that $B_c$ is equivalent to the average
number of centers found within an excluded volume, irrespectively of the spatial
configuration of the percolating objects.
However, for semi-penetrable spheres, the presence of hard-core introduces
a spatial correlation (see below), so that $B_c$ is expected to deviate from the uncorrelated
case of Eq. \eqref{eq:Bcspheres}. In particular, $B_c$ is found to decrease
as the hard-core portion of the sphere increases, reaching $B_c\simeq 1.5$ for
very thin penetrable shells \cite{Balberg1987,Heyes2006}, as a result of
the repulsion of the impenetrable hard-cores.

\begin{figure}[t]
  \begin{center}
  \includegraphics[scale=0.9, clip='true']{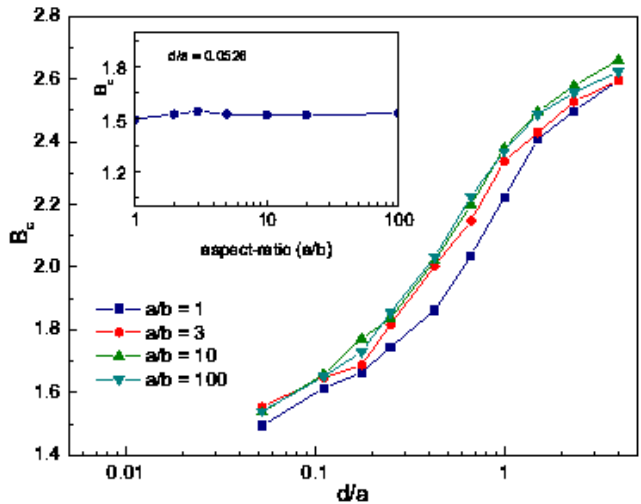}
  \caption{$B_{c}$ as a function of the shell thickness
  from simulation for different aspect-ratios. The results are obtained from the simulations by counting the connections number of each spheroid with its neighbors and averaging.}\label{fig:SEE3}
  \end{center}
\end{figure}

Let us now consider the case of assemblies of oblate ellipsoids. In Fig. \ref{fig:SEE3}
we plot the computed values of $B_c$  as a function of the penetrable shell thickness
$d/a$ and for selected values of the aspect-ratio $a/b$.
For $a/b=1$ we recover the results
for the spheres: $B_c\simeq 2.7$ for large values of $d/a$ while $B_c\simeq 1.5$ for
$d/a=0.0526$. For $a/b>1$ and thick penetrable shells, $B_c$ remains close to the spherical case
also for larger aspect-ratios because, as said before, for large $d/a$ values the
entire object (hard-core plus penetrable shell) is basically a semi-penetrable sphere
with a small hard-core spheroid. However, by decreasing $d/a$, we find that $B_c$ continues
to remain very close to the $a/b=1$ case also for the thinnest penetrable shells,
irrespectively of the aspect-ratio.
This is well illustrated by the inset of Fig. \ref{fig:SEE3} where the calculated
$B_c$ for $d/a=0.0526$ does not show appreciable variations over a
two-order of magnitude change of $a/b$. This result is rather interesting in view
of the fact that a quasi-invariance of $B_c$ with respect to the aspect-ratio in
oblate spheroids is in striking contrast to what is found in prolate objects
such as the spherocylinders studied in Refs.\cite{Balberg1987,Schilling2007}.
For example, for spherocylinders
made of hard cylinders on length $H$ and diameter $D$ capped by hemispheres and with
penetrable shells of thickness $0.1D$, $B_c$ is found to decreases from $B_c=1.61$ for
$H/D=4$ down to $B_c=1.29$ at $H/D=25$ \cite{Schilling2007}, consistently deviating
therefore from $B_c\simeq 1.76$ obtained for spheres of diameter $D$ and the same
penetrable shell thickness \cite{Heyes2006}. Different behaviors of quasi-impenetrable
oblate and prolate objects noted here are also found in the fully penetrable case.
Indeed $B_c$ of prolate objects decreases as the aspect-ratio is increased and
is expected to approach unity in the extreme prolate limit as a consequence
of the vanishing critical density \cite{Bug1985},
while $B_c$ of oblate objects remains close to $B_c\simeq 3$ all the way from the
moderate- to extreme-oblate regimes \cite{Garboczi1995}.

Now, we can write a general relation between the average connection number $B_c$
at percolation  and the critical number density $\rho_c$. If we consider hard spheroids
with penetrable shell and with an isotropic distribution of orientations, then $B_c$
reduces to:
\begin{equation}
\label{eq:Bgeneral}
B_c=\rho_c \int_{0}^{2\pi} \mathrm{d}\theta \int_{0}^{\pi}\mathrm{d}\varphi \,\Phi(\theta,\varphi)\int_{V_{exd}(\theta,\varphi)}\!\!\!\!\mathrm{d}^{3}\mathbf{r}\, g(\mathbf{r},\theta,\varphi),
\end{equation}
where $\theta$ and $\varphi$ are the angles between the major axes of two spheroids
separated by $\mathbf{r}$ and $g(\mathbf{r},\theta, \varphi)$ is the radial
distribution function: given a particle centered in the origin,
$\rho_c\,g(\mathbf{r},\theta, \varphi)$ represents the mean particle
number density at position $\mathbf{r}$ with an orientation
$\theta, \varphi$. The integration in $\mathbf{r}$ is performed over the
total excluded volume $V_{exd}(\theta,\varphi)$ (hard-core plus penetrable shell)
centered at the origin and having orientation $\theta,\varphi$.

We observe that all the information about the presence of a hard core
inside the particles is included in the radial distribution
function, which will be zero in the volume occupied by the hard
core of the particle centered at the origin. However, $g(\mathbf{r},\theta,\varphi)$ is a
rather complex function and even for the case of spheres there
are only approximate theoretical expressions
\cite{Trokhymchuk2005}. Also the construction of a fitted
expression to simulation data may result to be excessively complicated
when the respective orientation of the particles has to be taken
into account.

\begin{figure}[h!]
  \begin{center}
  \includegraphics[scale=0.5, clip='true']{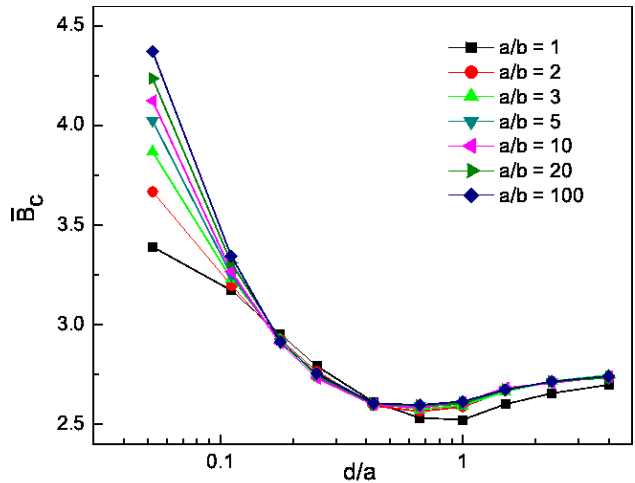}
  \caption{$\bar{B}_{c}$ as a function of the shell thickness for different aspect-ratios.}\label{fig:SEE4}
  \end{center}
\end{figure}

The lowest order approximation which we may then consider, and which is exact in
the case of fully penetrable particles, is the one where $g(\mathbf{r},\theta,\varphi)=1$.
This is equivalent to neglect all contributions of the radial distribution function
which spur from the presence of the hard core.
The resulting connectivity number, which in this approximation
we denote by $\bar{B}_c$, is then given by:
\begin{equation}
\label{Bbarra}
\bar{B}_c=\rho_c \int_{0}^{2\pi} \mathrm{d}\theta \int_{0}^{\pi}\mathrm{d}\varphi \,\Phi(\theta,\varphi)
\int_{V_{exd}(\theta,\varphi)}\!\!\!\!\mathrm{d}\mathbf{r}=\rho_c\langle V_{exd}\rangle,
\end{equation}
where $\langle V_{exd}\rangle$ is the orientation averaged total excluded volume.
Given the averaged excluded volume of spheroids
surrounded with a shell of constant thickness $\langle
V_{exd}\rangle$ (\ref{eq:<Vexd>isotr.}), together with the hard
spheroid excluded volume expression (\ref{eq:<Vex>isotr.}) or
(\ref{eq:<Vex>I.O.W..}), we can calculate $\bar{B}_c$ from the
percolation threshold results obtained from the simulations:
\begin{equation}
\label{Bbarra2}
\bar{B}_{c}=\rho_{c}\langle V_{exd}\rangle=\phi_{c}\frac{\langle V_{exd}\rangle}{V}\quad,
\end{equation}
where we have used the hard core volume fraction $\phi_c$. The full details of the
calculation of the excluded volume quantities can be found in the
appendix \ref{sec:excluded volume quantities}.
The resulting values of $\bar{B}_c$ are plotted in Fig. \ref{fig:SEE4} as
a function of the penetrable shell thickness and for several aspect-ratios.
Comparing Fig. \ref{fig:SEE4} with Fig. \ref{fig:SEE3} we note that for $d/a>1$,
i.e., for thick penetrable shells, $\bar{B}_c$ does not deviate much
from $B_c$, indicating that the effect of the hard-core is, in this regime,
rather weak. On the contrary, for thinner shells, $\bar{B}_c$
increasingly deviates from $B_c$ because the correlation driven by the hard-core
is stronger.

Despite that $\bar{B}_c$ overestimates the number of connected particles, its behavior is
nevertheless rather intriguing. Indeed the dependence of $\bar{B}_c$ on the penetrable
shell thickness $d/a$ appears to be universal with respect to the aspect-ratio, for all $d/a$ values
larger than $d/a>0.1$. Furthermore, in this region of $d/a$, $\bar{B}_c$ has a rather weak
dependence on the shell thickness, not deviating much from $\bar{B}_c\simeq 2.8$. This must be
contrasted to $B_c$ which, from $d/a=4$ down to $d/a=0.1$, decreases from about $2.8$ to only $1.6$.
The quasi-invariance of $\bar{B}_c$ has therefore some practical advantages since, by using
Eq. \eqref{Bbarra2}, the percolation threshold $\phi_c$ can be estimated from $\langle V_{exd}\rangle$
and $\bar{B}_c\simeq 2.8$, in a wide interval of $d/a$ and aspect-ratio values.

\section{Conclusions}

The geometrical percolation threshold in the continuum of random distributions
of oblate hard ellipsoids of revolution surrounded with a soft shell of constant
thickness has been investigated. Simulation results spanning a broad range of
aspect-ratios and shell thickness values have been reported. It is found that
larger aspect-ratios entail lower percolation thresholds, in agreement with
the behavior observed experimentally in insulator-conductor composites where
the conducting phase is constituted by oblate objects, such as graphite nanosheets.
Furthermore, the number $B_c$ of connected object at percolation is a quasi-invariant
with respect to the aspect-ratio, in contrast with what has been previously reported
for prolate objects. Finally, we have derived an additional quasi-invariant based on
the excluded volume concept which allows to infer the system percolation threshold.

\section{Acknowledgments}

This study was supported by the Swiss Commission for Technological Innovation (CTI)
through project GraPoly, (CTI grant 8597.2), a joint collaboration led by TIMCAL
Graphite \& Carbon SA. Simulations were performed at the ICIMSI facilities with
the help of Eric Jaminet. Data analysis was carried out with the help of Ermanno
Oberrauch. Figures \ref{fig:EV2} and \ref{fig:EV3} are due to Raffaele Ponti.
Comments by I. Balberg were greatly appreciated.

\appendix
\section{Evaluation of Excluded Volume Quantities}
\label{sec:excluded volume quantities}

\begin{figure}[t]
    \begin{center}
  \includegraphics[scale=0.4, clip='true']{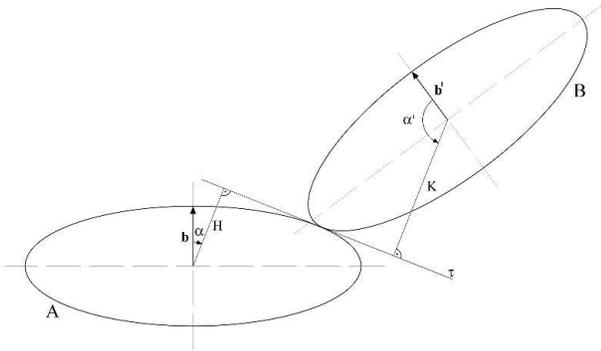}
  \caption{Two identical oblate spheroids in contact (2D representation).}\label{fig:EV2}
    \end{center}
\end{figure}

We report in the following the derivation of the excluded volume of two oblate spheroids, the excluded volume of two oblate spheroids surrounded with a shell of constant thickness and their angular averages. We follow a route due to the pioneering work of Isihara \cite{Isihara1951} which is somehow more laborious than the one used by the same author \cite{Isihara1950} and the authors of \cite{Ogston1975} to derive the widely used Isihara-Ogston-Winzor spheroid excluded volume formula. The advantage is that it is possible to obtain, albeit in a series expansion form, the excluded quantities with their full angle dependence. The average on the spheroid angle distribution function is performed in a second time and can be easily extended to non-isotropic cases.
Let us consider the case of two identical spheroids of major axis $a$ and minor axis $b$ in contact as illustrated in Fig. \ref{fig:EV2}.

We then have that the geometrical quantities $H$ and $K$, which represent the distances from the spheroid centers to the tangent plane to the two spheroids in the contact point, may be written as
\begin{equation}
\label{eq:H3}H(\alpha)=a\sqrt{1-\epsilon^2\cos^{2}{\alpha}},
\end{equation}
\begin{equation}
\label{eq:K1}
K(\alpha')=a\sqrt{1-\epsilon^2\cos^{2}{\alpha'}},
\end{equation}
where $\epsilon$ represents the \emph{eccentricity} (for oblate spheroids)
\begin{equation}
\label{eq:eccentricity}
\epsilon\equiv\sqrt{1-\frac{b^{2}}{a^{2}}}.
\end{equation}
Furthermore, we have
\begin{align}
\label{eq:cosalphaprime}
\cos^{2}{\alpha'}&=[\sin{\varphi}\sin{\alpha}(\cos{\theta}cos{\beta}
+\sin{\theta}\sin{\beta})+\cos{\varphi}\cos{\alpha}]^{2}\nonumber \\
&=[\sin{\varphi}\sin{\alpha}\cos{(\beta-\theta)}
+\cos{\varphi}\cos{\alpha}]^{2},
\end{align}
where $\theta$ and $\varphi$ are the angles which define the rotation that transforms the symmetry axis vector $\mathbf{b}$ of spheroid A in the one of B, $\mathbf{b}'$.

\begin{widetext}

We can then write the excluded volume of two identical spheroids, or more generally two identical ovaloids, as \cite{Isihara1951,Isihara1950}:
\begin{equation}
\label{eq:Vex1b}
V_{ex}=2V+\int K(H,H)\textrm{d}\omega=2V+\int_{0}^{2\pi}\textrm{d}\beta\int_{0}^{\pi}\textrm{d}\alpha\sin{\alpha}K(H,H),
\end{equation}
where $\textrm{d}\omega$ is the infinitesimal surface element of the unit sphere centered in the origin which, by using the reference frame choice of fig. \ref{fig:EV2}, takes the form
\begin{equation}
\label{eq:domega}
\textrm{d}\omega=\sin{\alpha}\textrm{d}\alpha\textrm{d}\beta.
\end{equation}
Furthermore, in Eq. (\ref{eq:Vex1b}) we have introduced the differential operator on the unit sphere which for two equal scalar quantities $F$ takes the form
\begin{equation}
\label{eq:diffop}
(F,F)\equiv2\Bigg\{\Bigg(\frac{\partial^{2}F}{\partial\alpha^{2}}+F\Bigg)\Bigg(\frac{1}{\sin^{2}{\alpha}}\frac{\partial^{2}F}{\partial\beta^{2}}+
\frac{\cos{\alpha}}{\sin{\alpha}}\frac{\partial F}{\partial\alpha}+F\Bigg)-\Bigg[\frac{\partial}{\partial\alpha}\Bigg(\frac{1}{\sin{\alpha}}\frac{\partial F}{\partial\beta}\Bigg)\Bigg]^{2}\Bigg\},
\end{equation}
while $V$ is the volume of the spheroid.

With the explicit form of $H$, Eq. (\ref{eq:H3}), $K$, Eq. (\ref{eq:K1}), and relation (\ref{eq:cosalphaprime}) we can write for the excluded volume (\ref{eq:Vex1b}) in the case of the two spheroids the integral form:
\begin{align}
\label{eq:Vex2}
V_{ex}(\theta,\varphi)&=2V+2a^{3}(1-\epsilon^{2})\int_{0}^{2\pi}
\textrm{d}\beta\int_{0}^{\pi}\textrm{d}\alpha\sin{\alpha}\frac{ \sqrt{1-\epsilon^{2}\cos^{2}{\alpha'}}}{(1-\epsilon^{2}\cos^{2}{\alpha})^{2}}\nonumber \\
&=2V+2a^{3}(1-\epsilon^{2})\underbrace{\int_{0}^{2\pi}\textrm{d}
\beta\int_{0}^{\pi}\textrm{d}\alpha\sin{\alpha}\frac{ \sqrt{1-\epsilon^{2}(\sin{\varphi}\sin{\alpha}\cos{\beta}+\cos{\varphi}\cos{\alpha})^{2}}}
{(1-\epsilon^{2}\cos^{2}{\alpha})^{2}}}_{I},
\end{align}
where we have used the fact that
\begin{equation}
\label{eq:notheta}
\int_{0}^{2\pi}\mathrm{d}\beta\sqrt{1-\epsilon^{2}[\sin{\varphi}\sin{\alpha}\cos{(\beta-\theta)}
+\cos{\varphi}\cos{\alpha}]^{2}} =\int_{0}^{2\pi}\mathrm{d}\beta\sqrt{1-\epsilon^{2}
(\sin{\varphi}\sin{\alpha}\cos{\beta}+\cos{\varphi}\cos{\alpha})^{2}},
\end{equation}
because of the $2\pi$ periodicity of the integrand and which means that $V_{ex}$ is $\theta$-independent.

We now may expand the $1-\epsilon^{2}(\sin{\varphi}\sin{\alpha}\cos{\beta}+\cos{\varphi}\cos{\alpha})^{2}$ square root:
\begin{align}
\label{eq:sqrt exp}
\sqrt{1-\epsilon^{2}(\sin{\varphi}\sin{\alpha}\cos{\beta}+\cos{\varphi}\cos{\alpha})^{2}}
&=1-\frac{1}{2\sqrt{\pi}}\sum_{k=1}^{\infty}\Gamma(k-\frac{1}{2})\frac{\epsilon^{2k}}{k!}
(\sin{\varphi}\sin{\alpha}\cos{\beta}+\cos{\varphi}\cos{\alpha})^{2k}\nonumber \\
&=1-\frac{1}{2\sqrt{\pi}}\sum_{k=1}^{\infty}\Gamma(k-\frac{1}{2})\frac{\epsilon^{2k}}{k!}
\sum_{i=0}^{k}\binom{k}{i}(\sin{\varphi}\sin{\alpha}\cos{\beta})^{2i}
(\cos{\varphi}\cos{\alpha})^{2k-2i}.
\end{align}
Substituting this in integral $I$ of (\ref{eq:Vex2}) and integrating in $\beta$ in the first
resulting term we obtain:
\begin{align}
\label{eq:int}
I=& 2\pi\int_{0}^{\pi}\textrm{d}\alpha\frac{\sin{\alpha}}{(1-\epsilon^{2}\cos^{2}{\alpha})^{2}}\nonumber \\
&-\frac{1}{2\sqrt{\pi}}\sum_{k=1}^{\infty}\Gamma(k-\frac{1}{2})\frac{\epsilon^{2k}}{k!}
\sum_{i=0}^{k}\binom{k}{i}\sin^{2i}{\varphi}\cos^{2k-2i}{\varphi}
\int_{0}^{2\pi}\textrm{d}\beta\cos^{2i}{\beta}\int_{0}^{\pi}\textrm{d}\alpha\frac{ \sin^{2i+1}{\alpha}\cos^{2k-2i}{\alpha}}{(1-\epsilon^{2}\cos^{2}{\alpha})^{2}}.
\end{align}
The integration follows then with the aid of formulas 2.153 (3.), 3.682, 3.681 (1.) of \cite{Gradshteyn2000} obtaining with (\ref{eq:Vex2}) the expression for the \emph{excluded volume of two identical oblate spheroids}:
\begin{align}
\label{eq:VexFinal}
V_{ex}(\varphi)=&2V+2a^{3}(1-\epsilon^{2})\biggl[4\pi F(\scriptstyle 2 \,,\,\frac{1}{2}\,,\,\frac{3}{2}\,,\,\epsilon^{2}\textstyle)\nonumber \\
&-\sqrt{\pi}\sum_{k=1}^{\infty}\Gamma( k-\frac{1}{2})\epsilon^{2k}
\sum_{i=0}^{k}\frac{\sin^{2i}{\varphi}\cos^{2k-2i}{\varphi}}{2^{i}(k-i)!(i!)^{2}}
B(\scriptstyle i+1\,,\,\frac{2k-2i+1}{2}\textstyle)
F(\scriptstyle 2\,,\,\frac{2k-2i+1}{2}\,,\,\frac{2k+3}{2}\,,\,\epsilon^{2}\textstyle)\biggr].
\end{align}

\begin{figure}[hhh]
    \begin{center}
  \includegraphics[scale=0.4, clip='true']{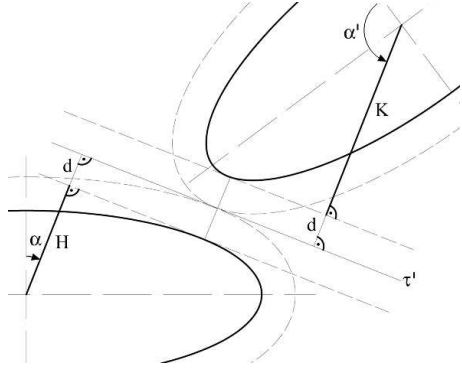}\\
  \caption{Two oblate spheroids surrounded with shells of constant thickness which are in contact (2D representation).}\label{fig:EV3}
    \end{center}
\end{figure}

Let us now consider the situation depicted in fig. \ref{fig:EV3} which represent two (identical) spheroids surrounded with a shell of constant thickness $d$. We are again interested in evaluating the excluded volume of these objects, which, because of the constant shell offset, will not be anymore ellipsoids. Nevertheless, we see that in this case we can again construct geometrical quantities like $H$ and $K$ of the two spheroids of Fig. \ref{fig:EV2} and that these, which we will call $H'$ and $K'$ are parallel to the old $H$ and $K$ respectively. Then it follows:
\begin{align}
\label{eq:Hprime}
H'(\alpha)&=H(\alpha)+d \\
K'(\alpha')&=K(\alpha')+d,
\end{align}
and $H$ and $K$ will be given by (\ref{eq:H3}) and (\ref{eq:K1}).
Now, also in this case expression (\ref{eq:Vex1b}) holds true and, observing that the volume of an ovaloid may be written as \cite{Isihara1950,Isihara1951}
\begin{equation}\label{eq:Vovaloid}V=\frac{1}{6}\int G(G,G)\mathrm{d}\omega,
\end{equation}
where $G$ is a geometric quantity constructed like $H, K, H', K'$, we are able to write for the excluded volume of the two spheroids with shell:
\begin{align}
\label{eq:Vexd}
V_{exd}=&2V'+\int K'(H',H')\mathrm{d}\omega\nonumber \\
=&V_{ex}+\frac{4d}{3}\underbrace{\int (H,H)\mathrm{d}\omega}_{I_{1}}+
2d\underbrace{\int\bigg(\frac{H}{3}+\frac{4d}{3}\bigg)
\bigg(\frac{\partial^{2}H}{\partial\alpha^{2}}+\frac{\cos{\alpha}}{\sin{\alpha}}\frac{\partial H}{\partial\alpha}+2H+d\bigg)\mathrm{d}\omega}_{I_{2}}\nonumber \\
&+2d\underbrace{\int K\bigg(\frac{\partial^{2}H}{\partial\alpha^{2}}
+\frac{\cos{\alpha}}{\sin{\alpha}}\frac{\partial H}{\partial\alpha}
+2H+d\bigg)\mathrm{d}\omega}_{I_{3}},
\end{align}
and $V_{ex}$ is the excluded volume of the two spheroids (\ref{eq:VexFinal}).

Integrals $I_{1}$ and $I_{2}$ are straightforward and may be solved with the aid of formulas 3.682, 2.583 (3.), 2.584 (3.) and 2.584 (39.) of \cite{Gradshteyn2000}:
\begin{equation}I_{1}=2a^{2}(1-\epsilon^{2})\int_{0}^{2\pi}\textrm{d}\beta\int_{0}^{\pi}\textrm{d}\alpha\frac{ \sin{\alpha}}{(1-\epsilon^{2}\cos^{2}{\alpha})^{2}}=8\pi a^{2}(1-\epsilon^{2})
F(\scriptstyle 2\,,\,\frac{1}{2}\,,\,\frac{3}{2}\,,\,\epsilon^{2}\textstyle),
\end{equation}

\begin{align}
\label{eq:I2b}
I_{2}&=\int_{0}^{2\pi}\textrm{d}\beta\int_{0}^{\pi}\textrm{d}
\alpha\sin{\alpha}\bigg(\frac{a\sqrt{1-\epsilon^{2}\cos^{2}{\alpha}}}{3}+\frac{4d}{3}\bigg)
\bigg[\frac{a}{\sqrt{1-\epsilon^{2}\cos^{2}{\alpha}}}+\frac{a(1-\epsilon^{2})}{(1-\epsilon^{2}\cos^{2}
{\alpha})^{\frac{3}{2}}}+d\bigg]\nonumber \\
&=\frac{4\pi}{3}(a^{2}+4d^{2})+6\pi ad\bigg(\sqrt{1-\epsilon^{2}}+\frac{\arcsin{\epsilon}}{\epsilon}\bigg)
+\frac{4\pi}{3}a^{2}(1-\epsilon^{2})\frac{\textrm{arctanh}\,\epsilon}{\epsilon}.
\end{align}
Regarding $I_{3}$ we have, using Eq. (\ref{eq:K1}), Eq. (\ref{eq:cosalphaprime}) and Eq. (\ref{eq:notheta}):
\begin{align}
I_{3}=&a\int_{0}^{2\pi}\!\!\!\textrm{d}\beta\int_{0}^{\pi}\!\!\textrm{d}\alpha
\sin{\alpha}\sqrt{1-\epsilon^{2}(\sin{\varphi}\sin{\alpha}
\cos{\beta}+\cos{\varphi}\cos{\alpha})^{2}}\nonumber \\
&\times\bigg[\frac{a}{\sqrt{1-\epsilon^{2}\cos^{2}{\alpha}}}+
\frac{a(1-\epsilon^{2})}{(1-\epsilon^{2}\cos^{2}
{\alpha})^{\frac{3}{2}}}+d\bigg],
\end{align}
and we can again expand the $1-\epsilon^{2}(\sin{\varphi}\sin{\alpha}\cos{\beta}+\cos{\varphi}\cos{\alpha})^{2}$ square root obtaining
\begin{align}
I_{3}=&a\int_{0}^{2\pi}\!\!\!\textrm{d}\beta\int_{0}^{\pi}\!\!\textrm{d}\alpha
\sin{\alpha}\bigg[\frac{a}{\sqrt{1-\epsilon^{2}\cos^{2}{\alpha}}}+
\frac{a(1-\epsilon^{2})}{(1-\epsilon^{2}\cos^{2}
{\alpha})^{\frac{3}{2}}}+d\bigg]\nonumber \\
&-\frac{a}{2\sqrt{\pi}}\sum_{k=1}^{\infty}\Gamma(k-\frac{1}{2})\frac{\epsilon^{2k}}{k!}
\sum_{i=0}^{k}\binom{k}{i}\sin^{2i}{\varphi}\cos^{2k-2i}{\varphi}
\int_{0}^{2\pi}\!\!\!\textrm{d}\beta\cos^{2i}{\beta}\nonumber \\
&\times\Bigg[a\!\int_{0}^{\pi}\!\!\mathrm{d}\alpha\frac{\sin^{2i+1}{\alpha}
\cos^{2k-2i}{\alpha}}{\sqrt{1-\epsilon^{2}\cos^{2}{\alpha}}}+
a(1-\epsilon^{2})\!\int_{0}^{\pi}\!\!\mathrm{d}\alpha
\frac{\sin^{2i+1}{\alpha}\cos^{2k-2i}{\alpha}}{(1-\epsilon^{2}\cos^{2}
{\alpha})^{\frac{3}{2}}}+d\!\int_{0}^{\pi}\!\!\mathrm{d}
\alpha\sin^{2i+1}{\alpha}\cos^{2k-2i}{\alpha}\Bigg].
\end{align}
These integrals may be solved again with the use of the formulas 2.153 (3.), 3.682, 3.681 (1.), 2.583 (3.), 2.584 (39.) and 3.621 (5.) of \cite{Gradshteyn2000}, yielding
\begin{align}
\label{eq:I3b}
I_{3}=&4\pi a\bigg(\frac{a\arcsin{\epsilon}}{\epsilon}+a\sqrt{1-\epsilon^{2}}+2d\bigg)
-a\sqrt{\pi}\sum_{k=1}^{\infty}\Gamma(k-\frac{1}{2})\epsilon^{2k}
\sum_{i=0}^{k}\frac{\sin^{2i}{\varphi}\cos^{2k-2i}{\varphi}}{2^{i}(k-i)!(i!)^{2}}\nonumber \\
&\times B(\scriptstyle i+1\,,\,\frac{2k-2i+1}{2}\textstyle)
\bigg[a F(\scriptstyle \frac{1}{2}\,,\,\frac{2k-2i+1}{2}\,,\,\frac{2k+3}{2}\,,\,\epsilon^{2}\textstyle)+a(1-\epsilon^{2})
F(\scriptstyle \frac{3}{2}\,,\,\frac{2k-2i+1}{2}\,,\,\frac{2k+3}{2}\,,\,\epsilon^{2}\textstyle)+d \bigg].
\end{align}
We can then combine all these results together with property \cite{Maple}
\begin{equation}F(\scriptstyle 2\,,\,\frac{1}{2}\,,\,\frac{3}{2}\,,\,\epsilon^{2}\textstyle)=
\frac{1}{2}\bigg(\frac{1}{1-\epsilon^{2}}+\frac{\textrm{arctanh}\,\epsilon}{\epsilon}\bigg)
\end{equation}
and Eq. (\ref{eq:Vexd}) to write the \emph{excluded volume of two oblate spheroids surrounded
with a shell of constant thickness} $V_{exd}$:
\begin{align}
\label{eq:VexdFinal}
V_{exd}=&V_{ex}+\frac{8\pi}{3} d(3a^{2}+4d^{2}+3ad)+4\pi ad(2a+3d)\bigg(\sqrt{1-\epsilon^{2}}+\frac{\arcsin{\epsilon}}{\epsilon}\bigg)\nonumber \\
&+8\pi a^{2}d(1-\epsilon^{2})\frac{\textrm{arctanh}\,\epsilon}{\epsilon}
-2ad\sqrt{\pi}\sum_{k=1}^{\infty}\Gamma{(k-\frac{1}{2})}\epsilon^{2k}
\sum_{i=0}^{k}\frac{\sin^{2i}{\varphi}\cos^{2k-2i}{\varphi}}{2^{i}(k-i)!(i!)^{2}}\nonumber \\
&\times B(\scriptstyle i+1\,,\,\frac{2k-2i+1}{2}\textstyle)
\bigg[a F(\scriptstyle \frac{1}{2}\,,\,\frac{2k-2i+1}{2}\,,\,\frac{2k+3}{2}\,,\,\epsilon^{2}\textstyle)+a(1-\epsilon^{2})
F(\scriptstyle \frac{3}{2}\,,\,\frac{2k-2i+1}{2}\,,\,\frac{2k+3}{2}\,,\,\epsilon^{2}\textstyle)+d \bigg].
\end{align}
We note that the above procedure allowed to obtain an expression for $V_{exd}$ with an angular dependence only upon $\varphi$. However, the orientation of the surface enclosing this volume will be dependent also on $\theta$, which is why it is needed e.g. in (\ref{eq:Bgeneral}).

The above results can also be easily used to compute the \emph{total volume of the spheroid with the shell} starting from Eq. (\ref{eq:Vovaloid}) with Eq. (\ref{eq:Hprime}):
\begin{equation}
V_{d}=V+\frac{d}{6}\int (H,H)\mathrm{d}\omega+
\frac{d}{3}\int(H+d)\bigg(\frac{\partial^{2}H}{\partial\alpha^{2}}+\frac{\cos{\alpha}}{\sin{\alpha}}\frac{\partial H}{\partial\alpha}+2H+d\bigg)\mathrm{d}\omega,
\end{equation}
which is very similar to the first part of Eq. (\ref{eq:Vexd}) and can be integrated alike, obtaining
\begin{equation}
\label{eq:Vd}
V_{d}=V+\frac{2\pi d}{3}\bigg[3a^{2}(1-\epsilon^{2})\frac{\textrm{arctanh}\,\epsilon}{\epsilon}+3ad\bigg(\sqrt{1-\epsilon^{2}}+
\frac{\arcsin{\epsilon}}{\epsilon}\bigg)+3a^{2}+2d^{2}\bigg].
\end{equation}

We now want to calculate the averaged excluded volume starting from the angle distribution functions which arise in the spheroid distributions of the simulation algorithm. For axially symmetric objects the angle distribution function $\Phi(\varphi)$ is dependent only on the angle between the symmetry axes $\varphi$. In the case of an isotropic (or Poissonian) angle distribution, where any orientation is equally probable, it is easy to find
\begin{equation}
\label{eq:PhiIsotr}
\Phi_{isotr.}(\varphi)=\frac{\sin{\varphi}}{4\pi}.
\end{equation}
To verify that this situation occurs unbiasedly in the simulations we used a modified version of the spheroid distribution creation algorithm where, after the distribution was realized, for every spheroid it was searched for neighbors which lied within a certain radius from its center and the angles between their symmetry axis were recorded. We then fitted the function to the simulated angle distribution results and, although we may expect that this distribution function will deviate from the purely isotropic case when highly packed assemblies are realized due to local orientation, we obtained no deviation for all binning radiuses and all volume fractions considered in the present research.
The averaged excluded volume of the two spheroids will then be
\begin{equation}
\langle V_{ex}\rangle_{isotr.}=\int_{0}^{2\pi}\mathrm{d}\theta\int_{0}^{\pi}\mathrm{d}\varphi\,
\Phi_{isotr.}(\varphi)V_{ex}(\varphi)=
\frac{1}{2}\int_{0}^{\pi}\mathrm{d}\varphi\,\sin{\varphi}V_{ex}(\varphi).
\end{equation}
This easily leads with (\ref{eq:VexFinal}) and 3.621 (5.) of \cite{Gradshteyn2000}
to the \emph{averaged excluded volume of two oblate spheroids}:
\begin{align}
\label{eq:<Vex>isotr.}
\langle V_{ex}\rangle=&2V+8\pi a^{3}(1-\epsilon^{2}) F(\scriptstyle 2 \,,\,\frac{1}{2}\,,\,\frac{3}{2}\,,\,\epsilon^{2}\textstyle)\nonumber \\
&-\sqrt{\pi}a^{3}(1-\epsilon^{2})\sum_{k=1}^{\infty}\Gamma( k-\frac{1}{2})\epsilon^{2k}
\sum_{i=0}^{k}\frac{[B(\scriptstyle i+1\,,\,\frac{2k-2i+1}{2}\textstyle)]^{2}}{2^{i}(k-i)!(i!)^{2}}
F(\scriptstyle 2\,,\,\frac{2k-2i+1}{2}\,,\,\frac{2k+3}{2}\,,\,\epsilon^{2}\textstyle),
\end{align}
and with Eq. (\ref{eq:VexdFinal}) and the same formula of \cite{Gradshteyn2000} to the \emph{averaged excluded volume of two oblate spheroids surrounded with a shell of constant thickness}:
\begin{align}
\label{eq:<Vexd>isotr.}
\langle V_{exd}\rangle=&\langle V_{ex}\rangle+\frac{8\pi}{3} d(3a^{2}+4d^{2}+3ad)+4\pi ad(2a+3d)\bigg(\sqrt{1-\epsilon^{2}}+\frac{\arcsin{\epsilon}}{\epsilon}\bigg)\nonumber \\
&+8\pi a^{2}d(1-\epsilon^{2})\frac{\textrm{arctanh}\,\epsilon}{\epsilon}
-ad\sqrt{\pi}\sum_{k=1}^{\infty}\Gamma{(k-\frac{1}{2})}\epsilon^{2k}
\sum_{i=0}^{k}\frac{[B(\scriptstyle i+1\,,\,\frac{2k-2i+1}{2}\textstyle)]^{2}}{2^{i}(k-i)!(i!)^{2}}\nonumber \\
&\times \bigg[a F(\scriptstyle \frac{1}{2}\,,\,\frac{2k-2i+1}{2}\,,\,\frac{2k+3}{2}\,,\,\epsilon^{2}\textstyle)+a(1-\epsilon^{2})
F(\scriptstyle \frac{3}{2}\,,\,\frac{2k-2i+1}{2}\,,\,\frac{2k+3}{2}\,,\,\epsilon^{2}\textstyle)+d \bigg].
\end{align}
The quantities involved in Eq. (\ref{eq:<Vex>isotr.}) and Eq. (\ref{eq:<Vexd>isotr.}) can then be easily evaluated with a mathematical software like Maple \cite{Maple} .

The averaged excluded volume of the hard spheroids (\ref{eq:<Vex>isotr.}) is of course equivalent to the Isihara-Ogston-Winzor expression \cite{Ogston1975,Isihara1950}:
\begin{equation}
\label{eq:<Vex>I.O.W..}
\langle V_{ex}\rangle_{I.O.W.}=\frac{4}{3}\pi a^{2}b\bigg\{2+\frac{3}{2}\bigg[1+\frac{\arcsin{\epsilon}}{\epsilon\sqrt{1-\epsilon^{2}}}\bigg]
\bigg[1+\frac{(1-\epsilon^{2})}{2\epsilon}\ln{\bigg(\frac{1+\epsilon}{1-\epsilon}\bigg)}\bigg]\bigg\}.
\end{equation}

These expressions have then been successfully verified through simulation by generating a great number of randomly placed spheroids couples with fixed reciprocal orientation and seeing how many times their shells overlapped. The ratio of overlaps to the total trial number will then be equal to the ratio of the excluded volume to the volume of the simulation cell. Convergence tests on the series were also performed.

It is then interesting to observe that the behavior ratio between $\langle V_{exd}\rangle$ and the spheroid volume $V$ is roughly linearly dependent upon the spheroid aspect-ratio and deviates slightly from it only close to the sphere case. The same holds true for the averaged excluded volume $\langle V_{ex}\rangle$, showing that interpreting the influence of the spheroid aspect-ratio as an excluded volume effect is a consistent approach.
\end{widetext}


\begin{thebibliography}{41}

\bibitem{Heyes2006} D. M. Heyes, M. Cass and A. C. Bra\'nca, Molec. Phys. {\bf 104}, 3137 (2006)
\bibitem{Lee1998} S. B. Lee and T. J. Yoon, J. Korean Phys. Soc. {\bf 33}, 612 (1998)
\bibitem{Wang1993} S. F. Wang and A. A. Ogale, Comp. Sci. Technol. {\bf 46}, 93 (1993)
\bibitem{Balberg1987} I. Balberg and N. Binenbaum, Phys. Rev. A {\bf 35}, 5174 (1987)
\bibitem{Shante1971} V. K. S. Shante and S. Kirkpatrik, Adv. Phys. {\bf 20}, 325 (1971)
\bibitem{Scher1970} H. Scher and R. Zallen, J. Chem. Phys. {\bf 53}, 3759 (1970)
\bibitem{Neda1999} Z. N\'eda, R. Florian, Y. Brechet, Phys. Rev. E {\bf 59}, 3717 (1999)
\bibitem{Bug1985} A. L. R. Bug, S.A. Safran and I. Webman, Phys. Rev. Lett. {\bf 54}, 1412 (1985)
\bibitem{Bug1986} A. L. R. Bug, S. A. Safran and I. Webman, Phys. Rev. B {\bf 33}, 4716 (1986)
\bibitem{Balberg1984a} I. Balberg, N. Binenbaum and N. Wagner, Phys. Rev. Lett. {\bf 52}, 1465 (1984)
\bibitem{Balberg1984b} I. Balberg, C. Anderson, S. Alexander and N. Wagner, Phys. Rev. B {\bf 30}, 3933 (1984)
\bibitem{Berhan2007b} L. Berhan and A. M. Sastry, Phys. Rev. E {\bf 75}, 041121 (2007)
\bibitem{Celzard1996} A. Celzard, E. McRae, C. Deleuse, M. Dufort, G. Furdin and J. F. Mar\^ech\'e, Phys. Rev. B {\bf 53}, 6209 (1996)
\bibitem{Charlaix1986} E. Charlaix, J. Phys. A: Math. Gen. {\bf 19}, L533 (1986)
\bibitem{Charlaix1984} E. Charlaix, E. Guyon and N. Rivier, Solid State Commun. {\bf 50}, 999 (1984)
\bibitem{Yi2004} Y. B. Yi and A. M. Sastry, Proc. R. Soc. Lond. A {\bf 460}, 2353 (2004)
\bibitem{Garboczi1995} E. J. Garboczi, K. A. Snyder, J. F. Douglas and M. F. Thorpe, Phys. Rev. E {\bf 52}, 819 (1995)
\bibitem{Sevick1988} E. M. Sevick, P. A. Monson and J. M. Ottino, Phys. Rev. A {\bf 38}, 5376 (1988)
\bibitem{Skal1974} A. S. Skal, B. I. Shklovskii, Sov. Phys. Semicond. {\bf 7}, 1058 (1974)
\bibitem{Berhan2007a} L. Berhan and A. M. Sastry, Phys. Rev. E {\bf 75}, 041120 (2007)
\bibitem{Ogale1993} A. A. Ogale and S. F. Wang, Comp. Sci. Technol. {\bf 46}, 379 (1993)
\bibitem{Schilling2007} T. Schilling, S. Jungblut and M. A. Miller, Phys. Rev. Lett. {\bf 98}, 108303 (2007)
\bibitem{Akagawa2007} S. Akagawa and T. Odagaki, Phys. Rev. E {\bf 76}, 051402 (2007)
\bibitem{Donev2005} A. Donev, S. Torquato and F. H. Stillinger, J. Comp. Phys. {\bf 202}, 765 (2005)
\bibitem{Perram1985} J. W. Perram and M. S. Wertheim, J. Comp. Phys. {\bf 58}, 409 (1985)
\bibitem{Vieillard1972} J. Vieillard-Baron, J. Chem. Phys. {\bf 56}, 4729 (1972)
\bibitem{Rimon1997} E. Rimon and S. P. Boyd, J. Intelligent Robotic Syst. {\bf 18}, 105 (1997)
\bibitem{Rimon1992} E. Rimon and S. P. Boyd, Tech. Rep. ISL, Stanford Univ. (1992)
\bibitem{Johner2008} N. Johner, C. Grimaldi, I. Balberg and P. Ryser, \emph{Accepted for pubblication on Phys. Rev. B}
\bibitem{Al-Futaisi2003} A. Al-Futaisi and T. W. Patzek, Physica A {\bf 321}, 665 (2003)
\bibitem{Hoshen1976} J. Hoshen and R. Kopelman, Phys. Rev. B {\bf 14}, 3438 (1976)
\bibitem{Rintoul1997} M. D. Rintoul and S. Torquato, J. Phys. A: Math. Gen. {\bf 30}, L585 (1997)
\bibitem{Lu2006} W. Lu, J. Weng, D. Wu, C. Wu and G. Chen, Mater. Manuf. Proc. {\bf 21}, 167 (2006)
\bibitem{Chen2003} G. Chen, C. Wu, W. Weng, D. Wu and W. Yan, Polymer {\bf 44}, 1781 (2003)
\bibitem{Stankovich2006} S. Stankovich, D. A. Dikin, G. H. B. Dommett, K. M. Kohlhaas, E. J. Zimney, E. A. Stach, R. D. Piner, S. T. Nguyen and R. S. Ruoff, Nature {\bf 442}, 282 (2006)
\bibitem{Trokhymchuk2005} A. Trokhymchuk, I. Nezbeda, J. Jirs\'ak and D. Henderson, J. Chem. Phys. {\bf 123}, 024501 (2005)
\bibitem{Isihara1951} A. Isihara, J. Chem. Phys. {\bf 19}, 1142 (1951)
\bibitem{Isihara1950} A. Isihara, J. Chem. Phys. {\bf 18}, 1446 (1950)
\bibitem{Ogston1975} A. G. Ogston and D. J. Winzor, J. Phys. Chem. {\bf 79}, 2496 (1975)
\bibitem{Gradshteyn2000} I. S. Gradshteyn and I. M. Ryzhik, \emph{Table of integrals, series, and products} (Academic Press, 2000)
\bibitem{Maple} Maple software by MapleSoft, a division of Waterloo Maple

\end{thebibliography}
\end{document}